# Turning-Point Impact Compromises the Biomechanical Integrity of Cortical Bone during Oscillating Sawing

Han Wang [a], Urara Satake [a, *], Toshiyuki Enomoto [a]

[a] *Division of Mechanical Engineering, Graduate School of Engineering, The University of Osaka, 2-1, Yamada-oka, Suita, Osaka, 565-0871, Japan*

*** Corresponding author:** Urara Satake [a]

Tel/ Fax: +81-6-6879-7287; E-mail: satake@mech.eng.osaka-u.ac.jp

**Statements and Declarations**

**Author Contributions:** All authors contributed to the study conception and design. Material preparation, data collection and analysis were performed by Han Wang, Urara Satake and Toshiyuki Enomoto. The first draft of the manuscript was written by Han Wang and all authors commented on previous versions of the manuscript. All authors read and approved the final manuscript.

**Funding:** This work was partially supported by JSPS KAKENHI Grant Numbers JP24K00777, JP25K23426.

**Data Availability:** The datasets and analysis files supporting the findings of this study are publicly available in Zenodo, DOI: 10.5281/zenodo.21919105.

The dataset will be made publicly available upon acceptance of the manuscript.

**Ethics approval:** The bovine femurs used in this study were obtained from a local commercial slaughterhouse as by-products of routine food processing. The animals were not sacrificed for research purposes. Therefore, ethical approval for animal experimentation was not required.

**Competing Interests:** The authors declare that they are inventors on a patent related to the optimized trajectory used in this study.

## Abstract

### Purpose

Oscillating bone sawing is widely used in orthopedic procedures such as total knee arthroplasty, where preservation of the mechanical integrity of the prepared bone surface is essential for implant stability. However, the reciprocating motion of oscillating saws can generate transient impact events near trajectory turning points, potentially inducing surface damage and reducing local load-bearing capacity. This study aimed to investigate the biomechanical consequences of turning-point impact loading during oscillating bone sawing.

### Methods

Oscillating sawing experiments were performed on bovine cortical bone using a conventional trajectory and an optimized trajectory designed to suppress turning-point impacts. Dynamic sawing forces were measured to quantify impact events. Surface damage on the groove wall surface was characterized using scanning electron microscopy (SEM) and optical imaging. Local and global mechanical responses were evaluated using compression tests.

### Results

The conventional trajectory produced pronounced periodic force spikes and generated extensive surface cracks and breakout regions. These impact-dominated events were associated with reduced local compressive strength of the prepared bone surface. In contrast, the optimized trajectory significantly suppressed peak impact forces, resulting in smoother groove wall surfaces with fewer defects. Consequently, the optimized trajectory increased the local compressive strength ($p$ < 0.001) and improved the global compressive strength of the specimens ($p$ = 0.001), while the elastic modulus remained largely unchanged.

**Conclusion**

Turning-point impacts during oscillating bone sawing compromise the biomechanical integrity of cortical bone by promoting surface damage and reducing local load-bearing capacity. Suppressing these transient impacts helps preserve the mechanical integrity of surgically prepared bone surfaces.

## Introduction

Oscillating saws are widely used in orthopedic procedures such as total knee arthroplasty (TKA) and high tibial osteotomy (HTO), where preservation of the mechanical integrity of the prepared bone surface is essential for postoperative stability and long-term implant fixation [1, 2]. Compared with rotary tools, oscillating saws offer better maneuverability and improved protection of surrounding soft tissues in confined anatomical regions [3]. However, the high-frequency interaction between the oscillating blade and heterogeneous cortical bone may generate transient mechanical disturbances that introduce microdamage to the prepared surface [4–6]. Because load transfer at the bone–implant interface is highly localized, even spatially confined weakening of implant-relevant regions may disproportionately affect interfacial stability and increase the risk of micromotion under physiological loading [7].

A key source of such localized damage is oscillation-induced impact occurring at the reciprocal turning points of the blade motion. Abrupt tooth re-engagement associated with non-uniform cutting depths can generate impact-dominant force components that substantially elevate the instantaneous load without directly contributing to material removal [8]. These impacts tend to concentrate near the oscillation turning points, where rapid directional changes trigger contact instability (Fig. 1a). Such repeated, high-magnitude loading induces subsurface cracks and edge defects that compromise the structural integrity of the bone surface. Furthermore, because these impacts recur at nearly identical spatial locations, damage inevitably accumulates locally rather than uniformly. Consequently, damaged and intact regions may coexist on the machined surface, producing spatial heterogeneity in local load-bearing capacity and potentially increasing the susceptibility of the bone–implant interface to micromotion (Fig. 1b).

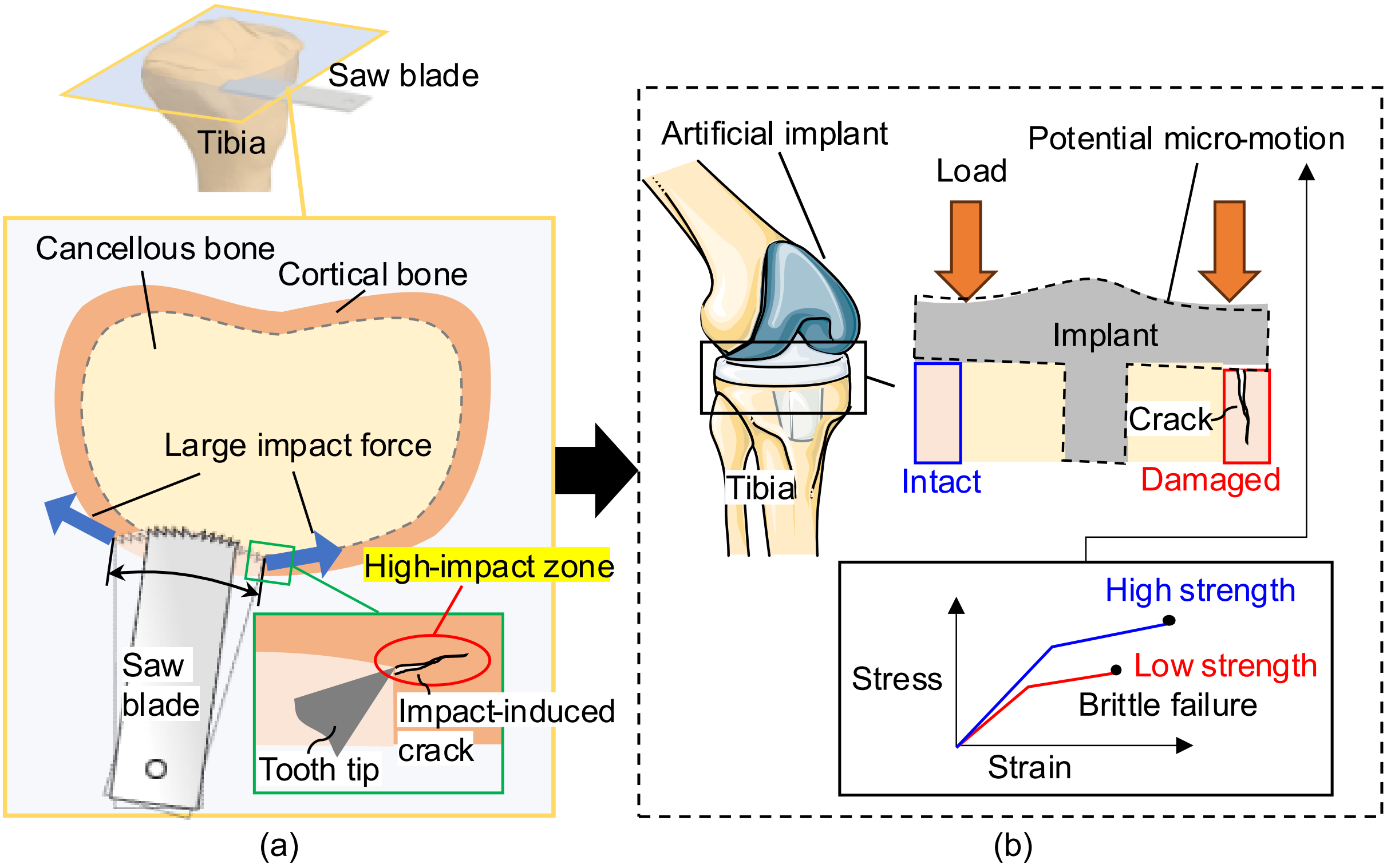

**Fig. 1** Conceptual relationship between oscillation-induced turning-point impacts and localized degradation of implant-relevant bone surfaces. (a) Cyclic impacts at reciprocal turning points may initiate subsurface cracks and edge defects near the turning-point region. (b) Under physiological loading, such localized damage can reduce the load-bearing capacity of the prepared surface and increase the susceptibility of the bone–implant interface to micromotion.

However, most previous evaluations of oscillating sawing performance have relied on global or averaged process-related metrics, including mean cutting forces [5, 6, 9], thermal injury [10–14], and surface topography [15–18]. While these metrics are useful for characterizing overall process performance, they are inherently insensitive to short-duration, high-magnitude impact events. As a result, the localized mechanical consequences of turning-point impacts may be underestimated. In particular, the reduction in mean cutting force does not necessarily imply preservation of the local load-bearing capacity of the prepared bone surface when impulsive mechanical disturbances are present. In our previous study, a trajectory-regulated condition was shown to effectively suppress turning-point impact severity [5]. However, whether such impact reduction can preserve the residual mechanical integrity of the affected bone surface remains unknown. This unresolved question highlights the need for localized mechanical evaluation of impact-damaged regions.

To evaluate such localized mechanical consequences, an appropriate mechanical testing scale is required. Conventional bulk compression tests assess global mechanical properties but inherently average out spatially confined defects. At the other extreme, micro- and nano-indentation techniques probe intrinsic tissue-level properties over very small material volumes and are strongly influenced by local microstructural features such as lamellar orientation and porosity [19, 20]. Although these methods are valuable for characterizing tissue-level properties, they do not necessarily represent the functional load-bearing capacity of implant-relevant bone surfaces, where larger-scale damage distribution and structural heterogeneity are more relevant. An intermediate-scale localized mechanical testing approach is therefore needed to bridge the gap between tissue-level characterization and macroscopic mechanical evaluation.

In this study, the effect of oscillation-induced impact on the mechanical integrity of prepared bone surfaces is investigated through a combined experimental framework. Dynamic cutting forces are first measured to identify the occurrence and magnitude of turning-point impacts under different oscillation trajectories. Surface morphology of the machined groove walls is then examined to determine how impact-induced damage propagates from the turning-point region toward the prepared surface. To quantify the mechanical consequences of this localized damage, a flat-punch compression method is employed to evaluate the regional load-bearing capacity of the machined surface, while global compression tests are conducted to assess the overall mechanical response of the sawn specimens. A conventional trajectory characterized by pronounced turning-point impacts is compared with a trajectory-regulated condition designed to suppress impact severity, so that the mechanistic role of turning-point impacts can be directly examined. By correlating cutting dynamics, surface damage features, and mechanical responses, this study elucidates how turning-point impact suppression influences the spatial distribution of residual load-bearing capacity on implant-relevant bone surfaces.

## Materials and Methods

The overall experimental workflow is shown in Fig. 2, including specimen preparation, oscillating sawing, surface damage observation, and compression testing.

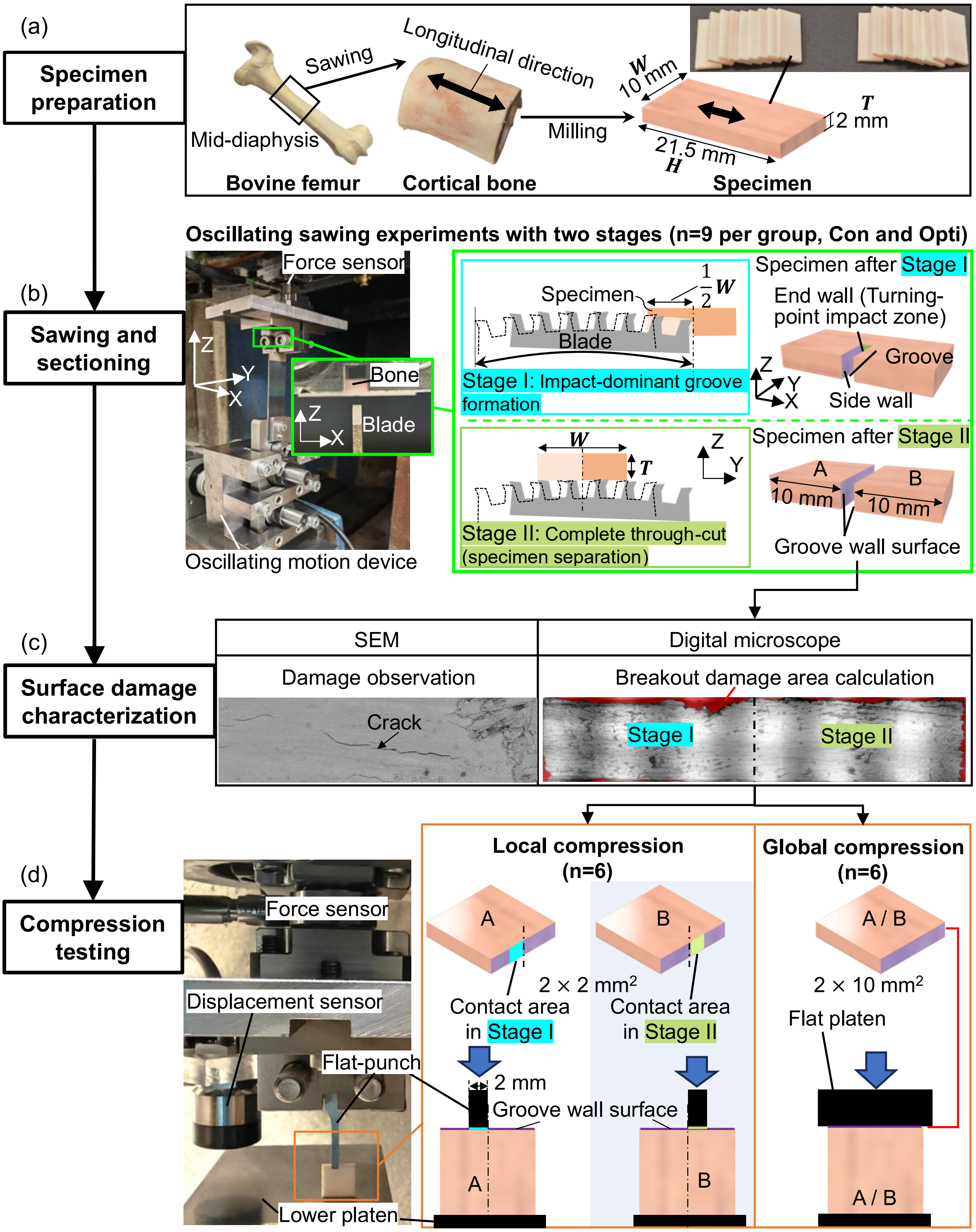

**Fig. 2** Experimental workflow of specimen preparation, oscillating sawing, surface

damage characterization, and compression testing. (a) Preparation of rectangular bovine cortical bone specimens from the mid-diaphysis of bovine femurs. (b) Oscillating sawing process and two-stage cutting protocol used to generate the machined surface and compression specimens: Stage I—impact-dominant groove formation near the turning point; Stage II—complete through-cut resulting in separated blocks. (c) Surface damage characterization by SEM for crack observation and digital microscopy for breakout damage area quantification. (d) Compression testing configuration, including local flat-punch compression on Stage I and Stage II regions and global compression of the entire specimen.

**Bone Specimen Preparation**

As shown in Fig. 2a, cortical bone blocks were extracted from the mid-diaphysis of a fresh adult bovine femur to minimize biological variability between specimens. The femur was obtained from a local commercial slaughterhouse as a by-product of the food industry. The blocks were precut into rectangular sawing specimens with dimensions of 10 mm (width, $W$) × 2 mm (thickness, $T$) × 21.5 mm (height, $H$), using a circular saw and a milling tool. These dimensions were selected to ensure stable oscillating sawing and to provide sufficient area for subsequent surface characterization and compression testing. To maintain structural consistency and mimic clinical sawing conditions [5], the height direction of each specimen was aligned parallel to the longitudinal osteon axis. After preparation, all specimens were wrapped in 0.9% physiological saline-soaked gauze, hermetically sealed in airtight plastic bags, and stored at −20 °C until use. The prepared specimens were used for oscillating sawing experiments, surface characterization, and subsequent mechanical testing. In addition, a small number of separate representative specimens were prepared specifically for scanning electron microscopy (SEM) observation to avoid any potential influence of vacuum exposure on the mechanical properties of specimens used in compression testing.

**Oscillating Sawing Experiments**

To evaluate the influence of oscillation-induced impact on bone surface integrity, sawing experiments were conducted under two motion conditions: a conventional oscillatory trajectory (Con) characterized by pronounced turning-point impacts, and an optimized trajectory (Opti) designed to suppress impact severity through trajectory regulation. Detailed mechanisms and structural parameters of the oscillating device have been reported previously [5]. Dynamic cutting forces were measured using a three-component force sensor (9327C, Kistler) at a sampling rate of 10 kHz. The force component along the oscillation direction (Y) was analyzed, and peak force events within each oscillation cycle were identified to quantify impact severity (Fig. 2b).

All experiments were performed at an oscillation frequency of 20 Hz and a feed speed of 0.1 mm/s, corresponding to an approximate cutting depth of 5 μm per oscillation cycle. These parameters were selected to ensure stable motion generation and repeatable force measurement under the present experimental platform.

A two-stage cutting protocol was implemented (Fig. 2b). In Stage I, an impact-dominant groove was generated using one side of the oscillating blade (blade thickness: 1.5 mm), so that the turning-point impact zone was formed at the groove end wall. The end wall was positioned at the mid-width of the specimen (5 mm from each side surface). In Stage II, cutting was continued along the same plane until through-cut separation was achieved, producing two compression blocks with dimensions of 10 mm $(W) \times$ 2 mm $(T) \times$ 10 mm $(H)$. In the study, the machined surface primarily refers to the groove wall surface generated during the sawing process, which serves as the main load-bearing surface for the subsequent mechanical evaluation. In Stage I, this surface includes both the side walls and the end wall of the groove. In Stage II, after through-cut separation, only the side wall surfaces remain.

Nine independent sawing experiments were conducted for each trajectory condition (Con and Opti; n = 9 per group). Since each sawing experiment yielded two compression blocks, a total of 36 blocks were obtained. For local flat-punch compression testing, 24 blocks were used, corresponding to n = 6 for each trajectory condition in each region (Con-Stage I, Con-Stage II, Opti-Stage I, and Opti-Stage II). The remaining 12 blocks were used for global compression testing, with n = 6 for each trajectory condition.

To minimize the influence of tool wear on cutting force measurement and surface damage formation, the condition of the blade teeth was inspected after each cut. The blade was replaced whenever visible tooth wear or deformation was observed before the subsequent experiment.

**Surface Damage Characterization**

Surface damage on the groove wall surface was characterized to evaluate trajectory-dependent crack and breakout formation (Fig. 2c). After sawing, the specimens were gently rinsed with saline solution and air-dried at room temperature.

Morphological features including microcracks, macrocracks, and local breakout defects were observed using a scanning electron microscope (TM3000, Hitachi). Representative regions from Stage I and Stage II portions of the groove wall surface were examined for qualitative comparison between the Con and Opti trajectories. To prevent potential alterations of bone mechanical properties caused by vacuum exposure, SEM observations were performed on separate representative specimens that were not included in the compression tests.

In addition, optical images of the groove wall surface were acquired using a digital microscope (VHX-EA300, Keyence). These images were analyzed using a Python-based program developed in this study. Breakout boundaries were manually annotated, and the remaining intact surface was automatically used to calculate the effective load-bearing area. This analysis provided both the global effective contact ratio of the groove wall surface, and the local effective contact area corresponding to the flat-punch compression test locations. The resulting effective area values were used for the corrected local compression analysis.

**Compression Testing**

To evaluate the mechanical consequences of sawing-induced surface damage, both local and global compression tests were performed using the separated bone blocks obtained after Stage II cutting (Fig. 2d). Compression loading was applied using the vertical translation axis of the machining-center-based platform at a constant displacement rate of 0.1 mm/s. The applied load was recorded using an inline force sensor (9327C, Kistler; maximum load: 8000 N), and specimen displacement was monitored using a non-contact eddy current displacement sensor (EX-422V, Keyence; resolution: 2 μm). System compliance was evaluated separately using a steel reference block. The measured machine displacement reached 0.062 mm at 6900 N, corresponding to approximately 20% of the total displacement observed during global specimen compression (~0.3 mm at peak strain). Because this displacement was not negligible relative to the specimen deformation, the corresponding compliance curve was used to correct the displacement data in the subsequent mechanical analysis [21]. To reduce friction between the specimen and the loading surfaces, a thin layer of liquid paraffin (mineral oil) was applied to both ends of each specimen prior to compression testing [22].

For the local compression test, a flat punch with a nominal contact area of 2 mm × 2 mm was used as the upper loading tool, while the specimen was supported on a lower platen. After Stage II cutting, each sawn specimen yielded two compression blocks corresponding to the Stage I and Stage II regions of the original groove wall surface, thereby forming a paired design. Local compression was performed at two symmetric positions defined relative to the block center. For the block corresponding to the Stage I region, the flat punch was positioned 1 mm to the left of the block center; for the block corresponding to the Stage II region, the flat punch was positioned 1 mm to the right of the block center (Fig. 2b,d). This arrangement ensured geometrically comparable loading conditions for the two regions. Local compression tests were conducted with n = 6 for each trajectory condition in each region (Con–Stage I, Con–Stage II, Opti–Stage I, and Opti–Stage II). To account for local surface breakout, the effective contact ratio was determined from the optical images of the groove wall surface (Fig. 2c), and the corresponding effective contact area was used for corrected local stress calculations.

For the global compression test, the same loading platform, lower platen, and measurement system were used, while the flat punch was replaced by a flat platen larger than the specimen cross section, so that unconfined compression was applied over the entire nominal cross section of the bone block. Compression was applied along the specimen height direction. The nominal cross-sectional area (10 mm × 2 mm) was used for the calculation of global compressive stress. Global compression tests were conducted with n = 6 for each trajectory condition.

Engineering stress–strain curves were calculated from the measured force and the compliance-corrected displacement. For local compression, both nominal stress and effective stress were calculated using the nominal contact area and the corrected effective contact area, respectively. For global compression, nominal stress was calculated using the nominal cross-sectional area. The peak stress prior to failure was defined as the compressive strength, and the corresponding strain was defined as the strain at peak stress. The elastic modulus was determined from the most linear segment of the stress–strain curve within a predefined low-strain region using a sliding-window linear regression procedure [23–25]. The detailed implementation of this procedure is provided in the accompanying MATLAB analysis code.

**Statistical Analysis**

All data are presented as mean ± standard deviation. Peak impact forces and local compression results were analyzed using a mixed two-way repeated-measures ANOVA, with trajectory condition (Con and Opti) as a between-subject factor and cutting stage/region (Stage I and Stage II) as a within-subject factor.

For the peak impact force analysis, Stage I and Stage II peak values were obtained sequentially from the same sawing trial, forming the repeated-measures factor. For the local compression tests, the two compression blocks corresponding to the Stage I and Stage II regions were obtained from the same parent specimen, forming a paired structure for the repeated-measures factor.

For the global compression tests, mechanical properties were compared between trajectory conditions using independent-samples t-tests.

Statistical significance was defined as $p < 0.05$. Normality and homogeneity of variance were verified prior to performing parametric tests. All statistical analyses were performed using MATLAB (MathWorks, USA).

## Results

### Impact Force Characteristics

The dynamic cutting force signals measured during oscillating bone sawing are shown in Fig. 3a–b. For the conventional trajectory, pronounced periodic force spikes were observed during Stage I (Fig. 3a), exhibiting substantially larger amplitudes than the baseline force fluctuations. In contrast, the optimized trajectory showed markedly smoother force signals during Stage I, with the large-amplitude spikes largely suppressed (Fig. 3b).

The statistical comparison of peak impact force is shown in Fig. 3c. During Stage I, the optimized trajectory produced a significantly lower peak impact force than the conventional trajectory ($p$ < 0.001). In contrast, no significant difference between the two trajectories was observed during Stage II. These stage-dependent differences are consistent with the two-way repeated-measures ANOVA results summarized in Table 1, which revealed significant effects of stage, trajectory, and their interaction (all $p$ < 0.001), indicating that the influence of trajectory on peak impact force varied between cutting stages.

**Table 1**

Two-way repeated-measures ANOVA results for peak force in the oscillation direction, with stage as a within-subject factor and trajectory as a between-subject factor.

| Source | $F(1,16)$ | $p$ | Partial $\eta^2$ |
|---|---|---|---|
| Stage | 832.5 | <0.001 | 0.981 |
| Trajectory | 338.6 | <0.001 | 0.955 |
| Trajectory × Stage | 1014.5 | <0.001 | 0.984 |

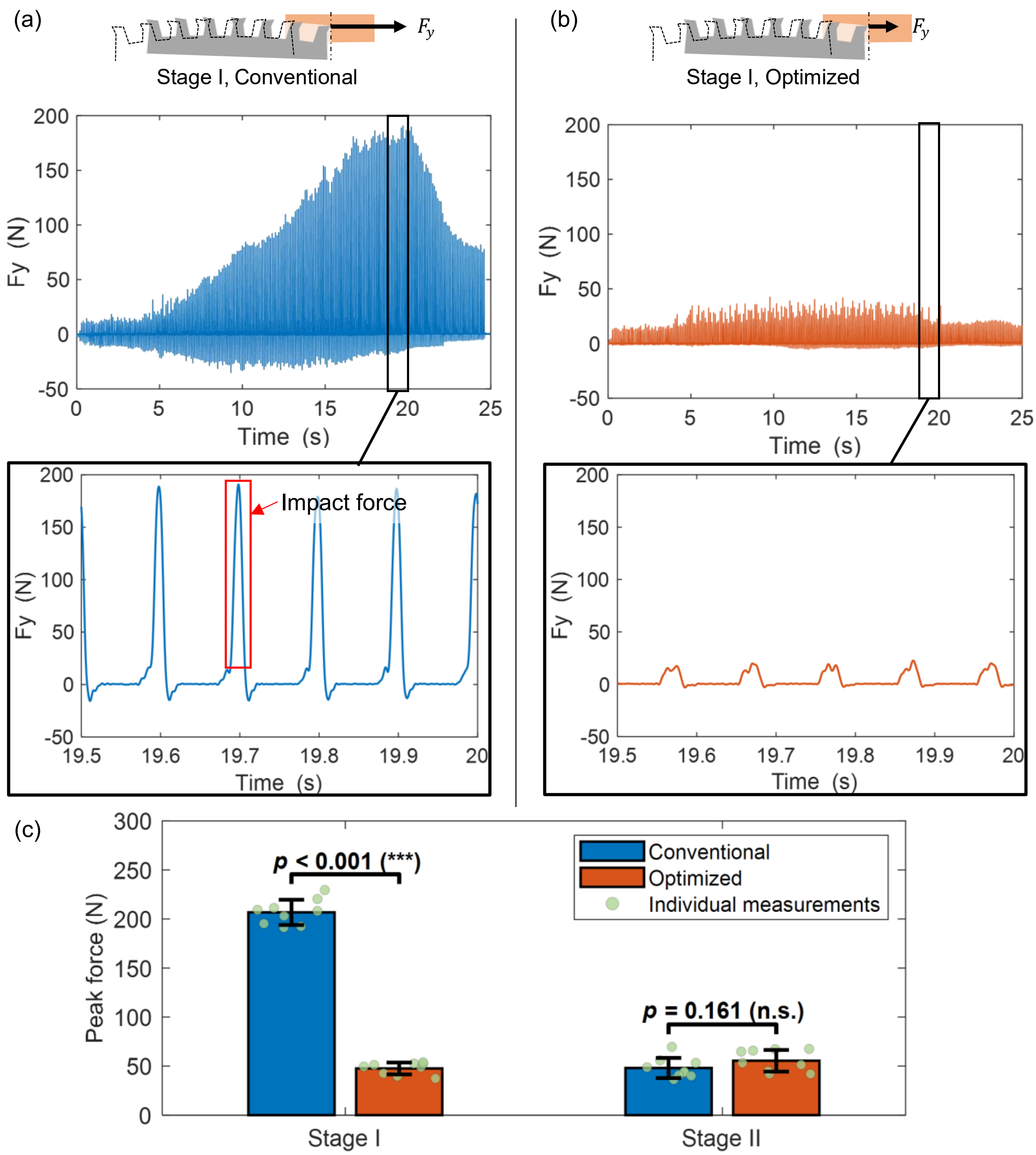

**Fig. 3** Dynamic impact characteristics of oscillating bone sawing under different trajectories during Stage I. Representative cutting force signal $F_y$ using (a) conventional and (b) optimized trajectories. (c) Statistical comparison of the peak impact force measured during Stage I and Stage II for the conventional and optimized trajectories.

**Surface Damage and Contact Area**

The surface morphology of the groove after Stage I was examined using scanning electron microscopy (SEM), as shown in Fig. 4. For the conventional trajectory, multiple cracks were observed on the groove wall surfaces (Fig. 4a–b). These cracks were mainly distributed near the end wall and along the side walls of the groove in the front view (Fig. 4a). In addition to surface cracks, localized material breakouts accompanied by macrocracks (defined as cracks longer than 300 μm [6, 24]) were observed at the groove edges in the top view (Fig. 4b).

In contrast, the surfaces produced using the optimized trajectory exhibited substantially fewer visible cracks (Fig. 4c–d). The groove wall surfaces appeared smoother in the front view (Fig. 4c), and large breakout regions were rarely observed in the top view (Fig. 4d).

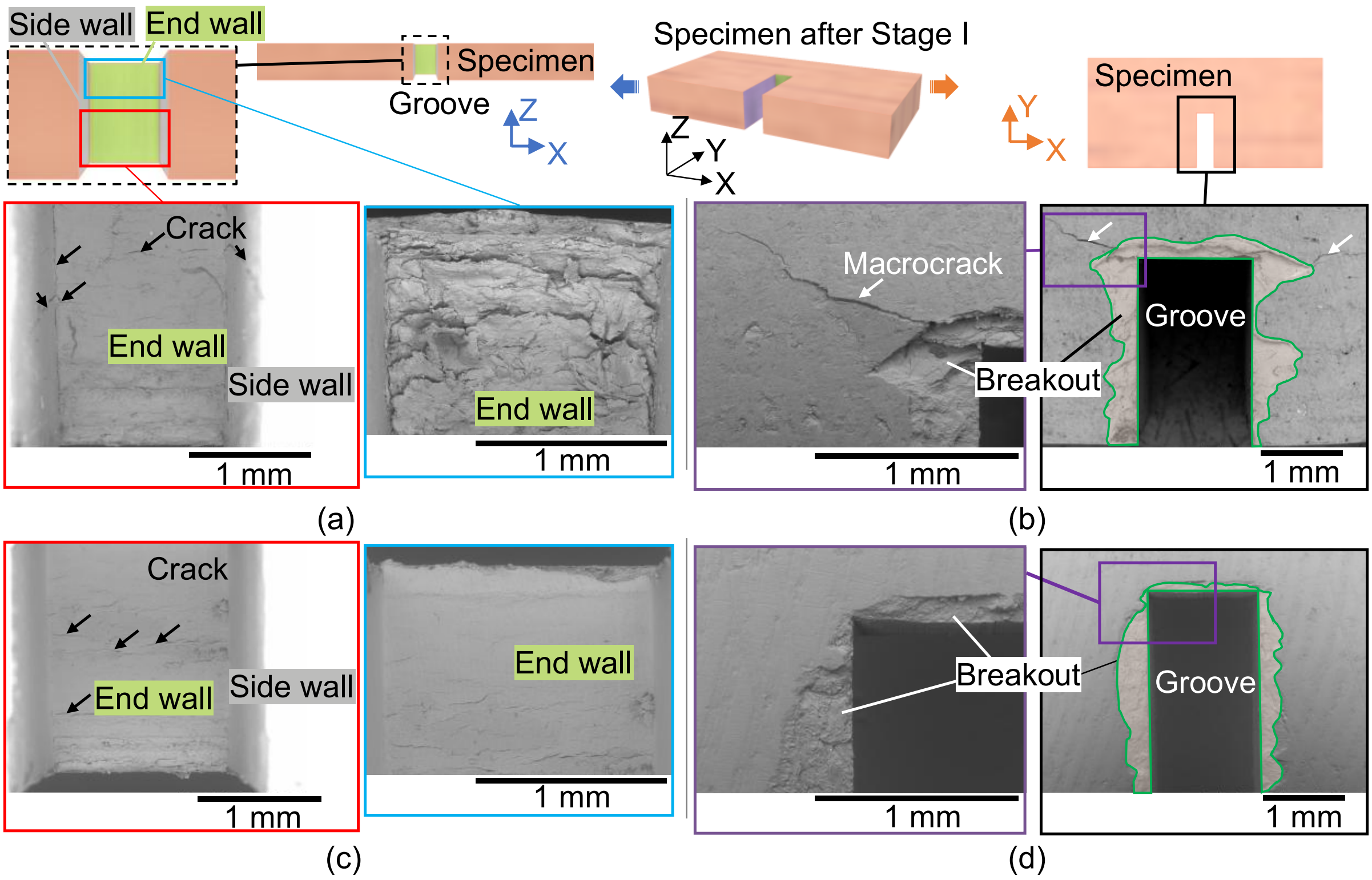

**Fig. 4** Surface morphology of the groove after Stage I cutting observed by SEM. Representative images obtained using different trajectories: (a) Conventional trajectory, front view; (b) Conventional trajectory, top view; (c) Optimized trajectory, front view; (d) Optimized trajectory, top view.

The surface damage features after Stage II cutting were examined by SEM (Fig. 5). For the conventional trajectory, distinct damage was observed on and adjacent to the groove wall surface. In the front view, the surface layer exhibited pronounced irregularities and visible ploughing marks, accompanied by both macrocracks and microcracks (Fig. 5a). Localized subsurface-damaged regions were also observed beneath the groove wall surface, with a representative depth of approximately 0.3 mm (Fig. 5a). In the top view, material breakout was observed along the groove edge in the XY plane, with a representative thickness of approximately 1.7 mm, and macrocracks were visible within the breakout area (Fig. 5b). The maximum breakout thickness was mainly concentrated near the central region of the groove wall surface.

For the optimized trajectory, the groove wall surface appeared more uniform in the front view, with fewer prominent surface irregularities (Fig. 5c). The damaged layer beneath the surface was thinner, with a representative depth of approximately 0.1 mm (Fig. 5c). In the top view, breakout features were still observed, but the affected region was smaller, with a representative thickness of approximately 0.4 mm, and only microcracks were observed within the breakout area (Fig. 5d). In contrast to the conventional trajectory, the breakout regions were more dispersed along the groove edge, and the maximum breakout thickness did not show a clear concentration near the center.

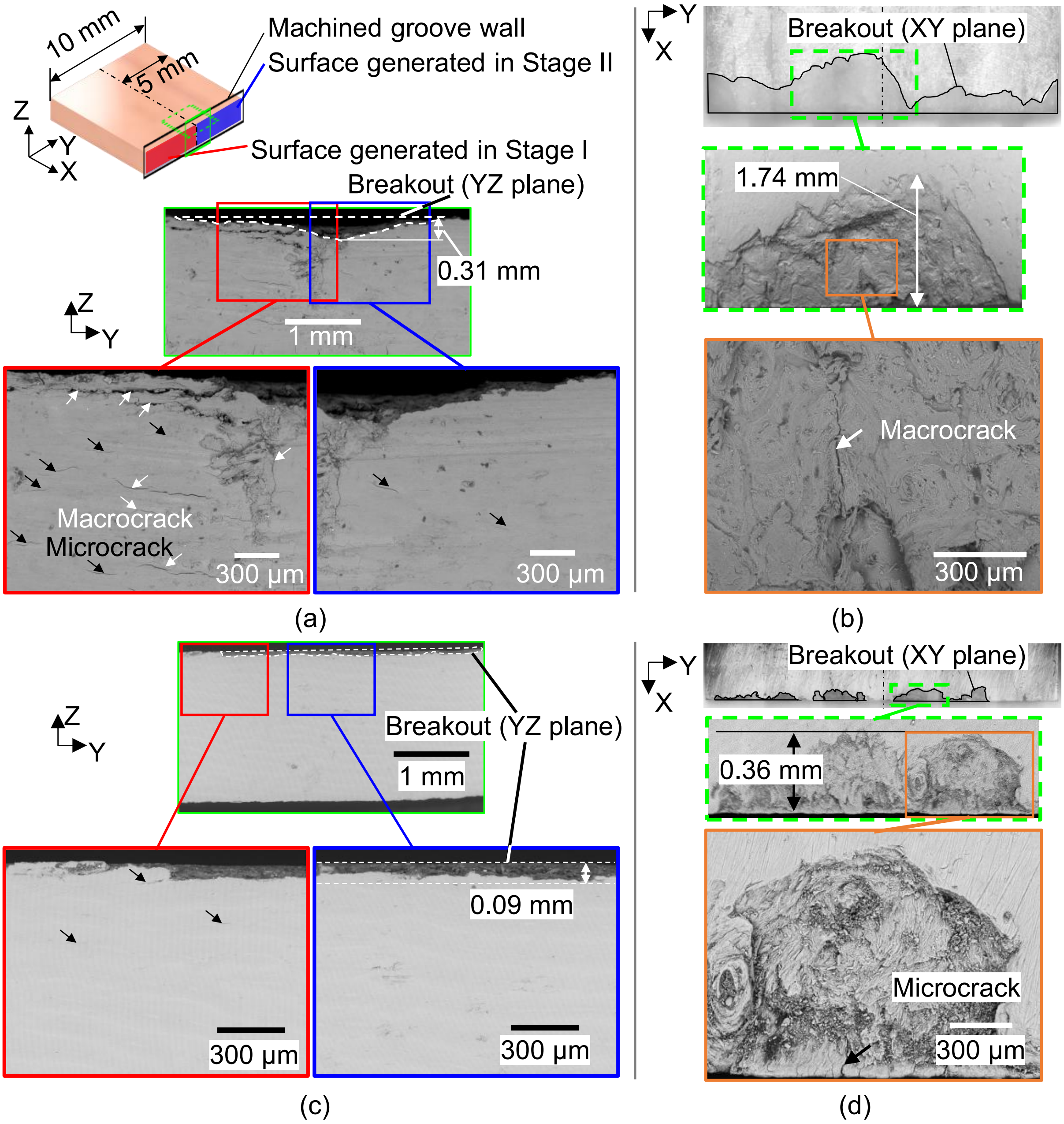

**Fig. 5** Surface damage features on the groove wall surface after Stage II cutting. Representative SEM observations for the conventional and optimized trajectories from different viewing directions: (a) conventional trajectory, front view; (b) conventional trajectory, top view; (c) optimized trajectory, front view; (d) optimized trajectory, top view.

To quantitatively evaluate the influence of cutting trajectory on surface integrity, the effective contact area was measured, as shown in Fig. 6. For the conventional trajectory, the global effective contact ratio was relatively low (Fig. 6a), indicating that a considerable portion of the groove wall surface was affected by surface damage. The local measurements revealed a clear difference between the two regions (Fig. 6b). In the conventional trajectory, the effective contact ratio in the Stage I region was approximately 92.8%, which was lower than that in the Stage II region (approximately 96.0%). In contrast, the optimized trajectory produced higher effective contact ratios in both regions, with values of approximately 96.7% in Stage I and 97.5% in Stage II. These results indicate that the optimized trajectory preserved a larger effective load-bearing area on the groove wall surface, particularly in the Stage I region where turning-point impact was more pronounced.

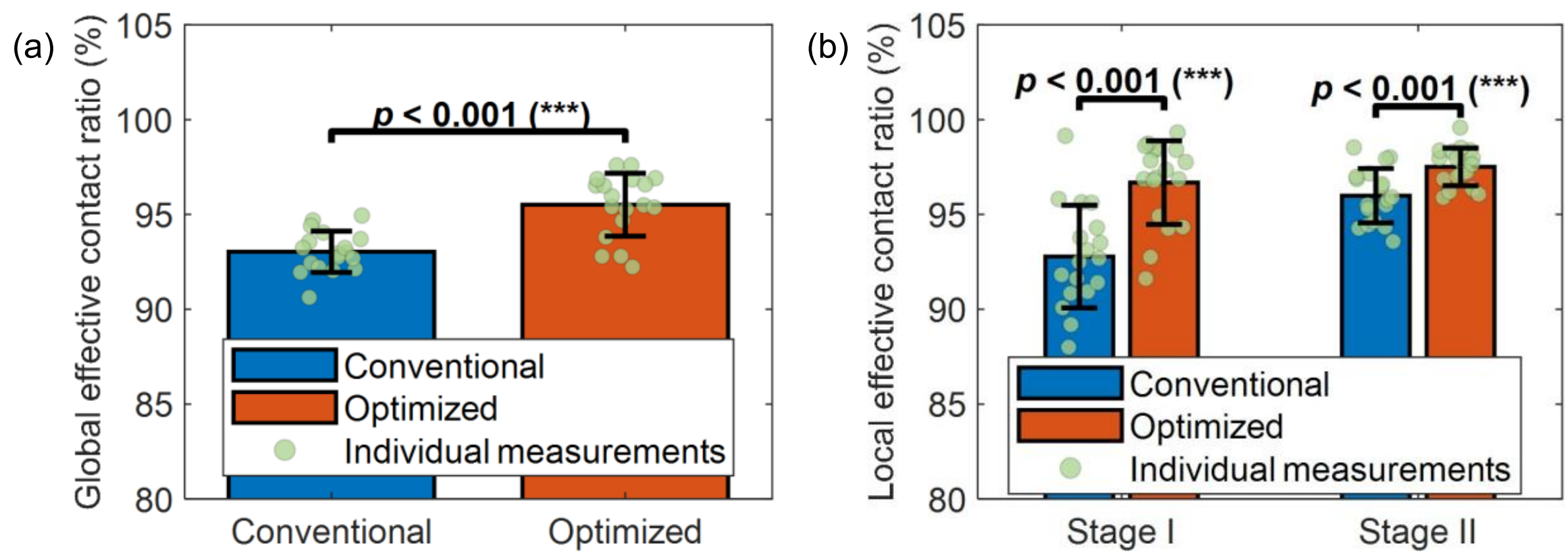

**Fig. 6** Quantification of effective contact area on the groove wall suraces. (a) Global effective surface ratio measured for the conventional and optimized trajectories. (b) Local effective contact ratio measured in the Stage I and Stage II regions of the machined surface.

**Local Mechanical Response**

Local compression tests were conducted on compression blocks corresponding to the Stage I and Stage II regions of the groove wall surface. The nominal mechanical responses obtained under the conventional trajectory are shown in Fig. 7a. Under the conventional trajectory, Stage II specimens exhibited higher peak stresses than Stage I specimens, and the compressive strength of Stage II was significantly higher than that of Stage I ($p$ < 0.001). In contrast, no statistically significant Stage I–Stage II differences were observed in elastic modulus ($p$ = 0.247) or strain at peak stress ($p$ = 0.773) .

The corresponding nominal responses obtained under the optimized trajectory are presented in Fig. 7b. Compared with the conventional trajectory, the responses of the Stage I and Stage II regions were closer to each other, and no significant Stage I–Stage II differences were detected in compressive strength, elastic modulus, or strain at peak stress.

To account for the reduction in load-bearing area caused by visible damaged regions on the groove wall surface, the local mechanical properties obtained under the conventional trajectory were recalculated using the effective contact area. The comparison between nominal and corrected mechanical properties is shown in Fig. 7c. After effective-area correction, the Stage I–Stage II difference in compressive strength decreased from $\Delta_{nom}$ = 27.9 MPa to $\Delta_{eff}$ = 16.7 MPa, while remaining statistically significant ($p_{eff}$ = 0.019). For elastic modulus, the Stage I–Stage II difference decreased from $\Delta_{nom}$ = 4.3 GPa to $\Delta_{eff}$ = 3.0 GPa and remained statistically insignificant ($p_{eff}$ = 0.400).

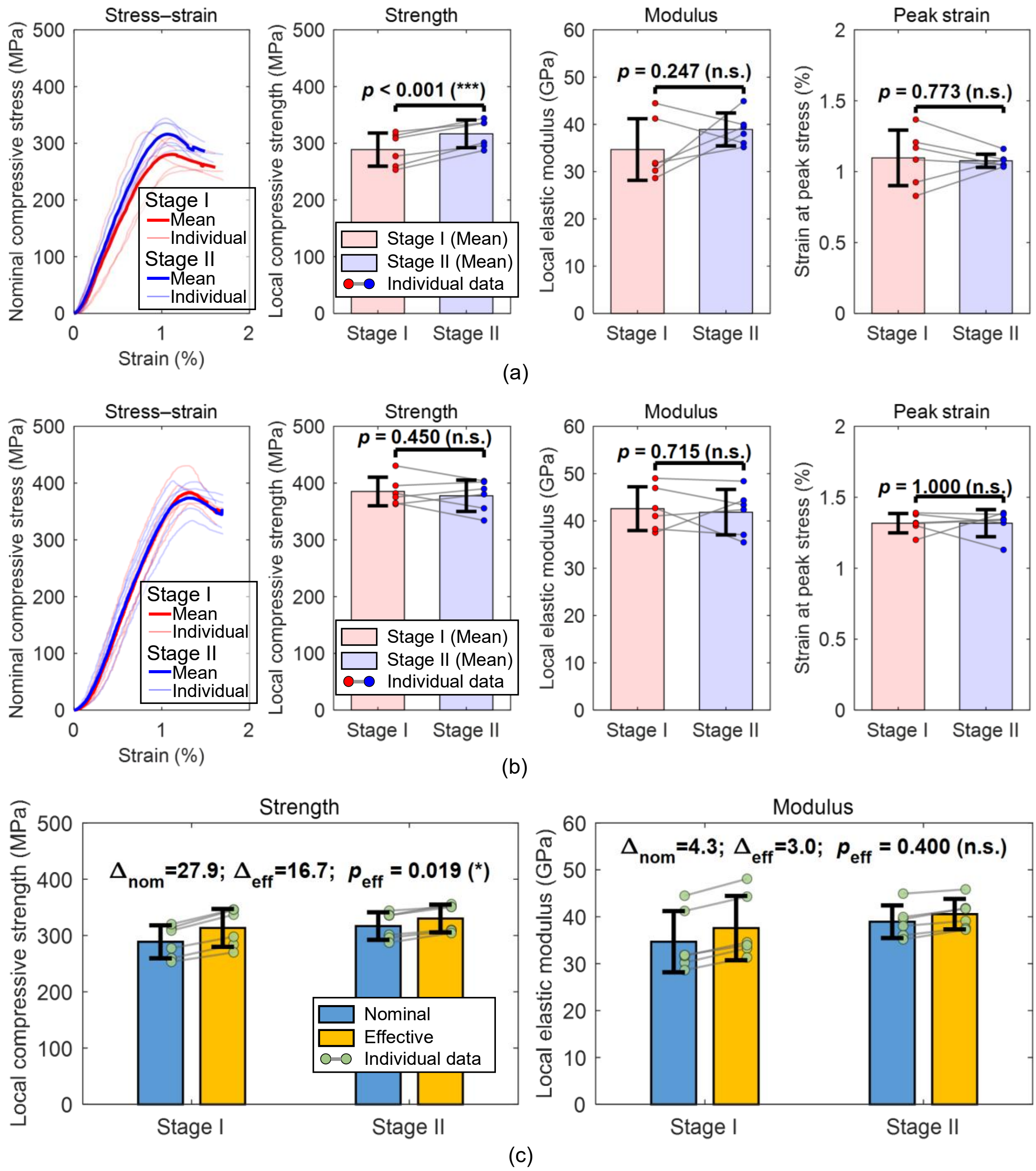

**Fig. 7** Local mechanical response of bone regions generated during oscillating sawing. Nominal compression results for specimens extracted from Stage I and Stage II regions obtained using (a) conventional (b) optimized trajectories. (c) Comparison between nominal and effective mechanical properties for the conventional trajectory.

The two-way repeated-measures ANOVA results for local compressive strength are summarized in Table 2. A significant main effect of trajectory was observed ($p <$ 0.001), indicating that the optimized trajectory generally produced higher local compressive strength than the conventional trajectory. The main effect of region did not reach statistical significance ($p$ = 0.071), although Stage II specimens tended to exhibit slightly higher strength than Stage I specimens. Importantly, a significant trajectory × region interaction was identified ($p$ = 0.005), indicating that the effect of trajectory on local compressive strength depended on the region of the groove wall surface.

**Table 2**
Two-way repeated-measures ANOVA results for local compressive strength, with region as a within-subject factor and trajectory as a between-subject factor.

| Source | $F(1,10)$ | $p$ | Partial $\eta^2$ |
|---|---|---|---|
| Trajectory | 29.14 | <0.001 | 0.745 |
| Region | 4.09 | 0.071 | 0.290 |
| Trajectory × Region | 12.43 | 0.005 | 0.554 |

**Global Mechanical Response**

The global compressive response of bone specimens processed using the conventional and optimized trajectories is shown in Fig. 8. Representative stress–strain curves indicate that specimens cut using the optimized trajectory generally exhibited higher peak stresses than those produced using the conventional trajectory. This trend is reflected in the measured compressive strength, which was significantly higher for the optimized trajectory than for the conventional trajectory ($p$ = 0.001). In contrast, the elastic modulus showed no statistically significant difference between the two trajectories ($p$ = 0.121). Similarly, the strain at peak stress remained comparable between the two groups ($p$ = 0.626). Overall, the optimized trajectory improved the global compressive strength while maintaining similar elastic modulus and peak strain compared with the conventional trajectory, consistent with the improved surface integrity observed at the groove wall surface.

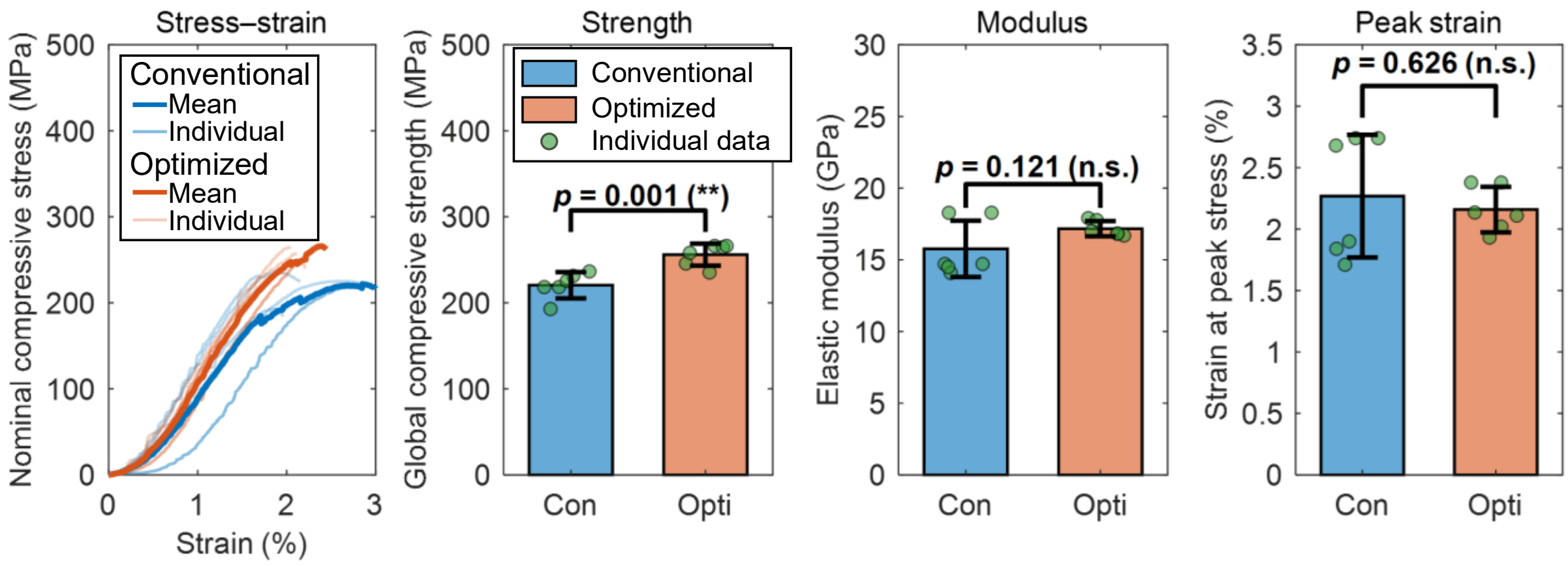

**Fig. 8** Global compression response of bone specimens after oscillating sawing with the conventional and optimized trajectories.

## Discussion

### Turning-Point Impact Damage

The present results indicate that the conventional oscillatory trajectory generates periodic impact-dominant loading at the reciprocating turning points, leading to localized damage on the groove wall surface. This interpretation is supported by the dynamic force measurements (Fig. 3), in which the conventional trajectory exhibited pronounced periodic force spikes during Stage I, whereas the optimized trajectory markedly suppressed these high-amplitude transients. These events represent non-steady-state load peaks that increase the instantaneous contact stress without proportionally enhancing material removal, thereby increasing the likelihood of crack initiation and near-surface damage.

Consistent with the force characteristics, the intermediate-stage morphology after Stage I directly captured the early damage features. Under the conventional trajectory, multiple cracks were observed on the groove wall surfaces (Fig. 4a–b), mainly distributed near the end wall and extending toward adjacent regions along the side walls of the groove. The coexistence of surface cracking and localized breakout accompanied by macrocracks in the top view (Fig. 4b) suggests that turning-point impacts can serve as crack nucleation sites and promote crack growth beyond the immediate impact-contact zone through stress redistribution within the groove geometry. Further support for this mechanism was obtained after complete sectioning of the bone specimens in Stage II. Severe breakout regions and macrocracks were observed in areas that had been subjected to repeated impact exposure (Fig. 5a–b), indicating that the damage was not limited to the initially impacted location but developed into more extensive surface and subsurface deterioration.

In contrast, the surfaces produced using the optimized trajectory exhibited fewer cracks and smoother groove wall surfaces (Fig. 4c–d), and the damaged layer beneath the groove wall surface was substantially thinner (Fig. 5c–d). The reductions in breakout size, crack density, and subsurface damage depth indicate that suppressing the turning-point impact effectively inhibits both the initiation and propagation of fractures in the near-surface region.

The local mechanical results further support this interpretation. As shown in Fig. 7a, specimens corresponding to the Stage I region, where impact exposure was more pronounced, exhibited significantly lower compressive strength than those corresponding to the Stage II region under the conventional trajectory ($p$ < 0.001). In contrast, the elastic modulus and strain at peak stress remained statistically unchanged. This pattern suggests that impact-induced damage primarily introduces microstructural defects that reduce failure strength without substantially altering elastic stiffness. Such behavior is consistent with microcrack-mediated damage in cortical bone, where microcrack formation acts as an early damage mechanism that degrades strength while the overall elastic stiffness remains largely preserved during the initial stages of damage accumulation [26–28]. These findings extend our previous trajectory-based observations [5] by showing that turning-point impacts damage the groove corners and end-wall regions, as well as deteriorate the load-bearing groove wall surface in a stage-dependent manner. Those results consistent with the crack-extension tendency reported under elevated thrust/ploughing conditions in our previous tool-structure study [6], suggesting that distinct loading modes may converge in promoting subsurface fracture growth along the groove wall surface.

### Region-Dependent Mechanical Degradation

Turning-point impact damage produced clear spatial heterogeneity in the mechanical integrity of the groove wall surface. This was first reflected in the effective contact area measurements: under the conventional trajectory, the Stage I region exhibited a lower effective contact ratio than the Stage II region, whereas the optimized trajectory produced higher and more uniform values across both regions (Fig. 6). These results indicate that the conventional trajectory caused more severe surface loss and breakout in the early impact-exposed region.

The local compression results showed that this spatial variation in surface condition translated into region-dependent mechanical performance. Under the conventional trajectory, the Stage II region exhibited significantly higher compressive strength than the Stage I region (Fig. 7a, $p < 0.001$), whereas under the optimized trajectory the Stage I–Stage II difference became much smaller and was no longer statistically significant (Fig. 7b, $p = 0.450$). After correction using the effective load-bearing area, the Stage I–Stage II strength difference decreased but remained significant (Fig. 7c), indicating that part of the apparent strength reduction originated from the loss of load-bearing area caused by surface damage, while the remaining difference suggests additional subsurface mechanical degradation. In contrast, the elastic modulus remained statistically unchanged before and after correction, indicating that the dominant effect of the damage was on failure-related resistance rather than on the initial elastic response.

The two-way repeated-measures ANOVA results (Table 2) further support this interpretation. A significant main effect of trajectory was observed ($p < 0.001$), indicating that the optimized trajectory improved the local compressive strength of the groove wall surface overall. The main effect of region did not reach statistical significance ($p = 0.071$), although the Stage II region tended to exhibit slightly higher strength than the Stage I region. More importantly, the significant trajectory × region interaction ($p = 0.005$) indicates that the mechanical benefit of trajectory optimization was region-dependent and was more pronounced in the impact-dominant Stage I region. This finding suggests that suppressing turning-point impact improves local strength as well as reduces spatial heterogeneity in the mechanical integrity of the groove wall surface.

**Implant-Relevant Surface Integrity**

These findings have direct implications for the integrity of surgically prepared, implant-relevant bone surfaces. Under the conventional trajectory, repeated turning-point impacts reduced the effective contact area and lowered the local compressive strength of the groove wall surface, indicating a reduction in local load-bearing capacity. Although the elastic modulus remained largely unchanged, the decrease in strength suggests that the damaged regions may act as preferential sites for failure initiation under subsequent loading. By contrast, the optimized trajectory produced fewer surface defects and a more uniform local mechanical response, indicating that suppressing turning-point impacts helps preserve the biomechanical integrity of the surgically prepared bone surface (Figs. 4–7).

The global compression results further support this interpretation. Specimens processed using the optimized trajectory exhibited higher compressive strength than those processed using the conventional trajectory, while the elastic modulus and strain at peak stress remained comparable (Fig. 8). This trend is consistent with the local compression results and suggests that localized surface damage induced by turning-point impacts can accumulate and ultimately reduce the overall load-bearing capacity of the specimen. In this sense, suppressing transient impact events not only improves the local integrity of the groove wall surface, but also contributes to better global mechanical performance of the prepared bone.

### Surgical Stability and Monitoring

The present findings also have implications for surgical stability and real-time monitoring. Because turning-point impacts appear as distinct spikes in the cutting-force signal, they may serve as practical indicators of unstable cutting conditions. Repeated high-amplitude spikes may reflect an increased risk of crack formation and surface breakout, whereas smoother signals with fewer transients may indicate a more stable cutting state with better surface integrity.

From a surgical technology perspective, the cutting-force signal could therefore be used as a feedback variable for monitoring oscillating bone sawing in real time. In practical settings, transient impact events may contribute to several instability-related phenomena. Asymmetric impact loading may introduce lateral force imbalance, potentially leading to unintended blade deviation or “kicking-out” behavior [29, 30], which may increase the risk of contact with surrounding soft tissues or unintended damage outside the planned cutting plane [32]. Repeated impact loading may also increase cutting resistance, thereby contributing to blade grabbing or temporary motor stalling, phenomena occasionally reported in clinical practice [31, 33]. Although these phenomena were not directly investigated in the present study, the suppression of transient force spikes under the optimized trajectory suggests that trajectory regulation may help mitigate instability-related risks during oscillating bone cutting.

### Limitations and Prospects

Several limitations of the present study should be acknowledged. First, the experiments were conducted under controlled laboratory conditions using bovine cortical bone extracted from the same femur to minimize biological variability. While this approach improves experimental consistency, it does not fully represent the anatomical and mechanical variability of human bone encountered clinically. Second, although the effective-area correction indicated that part of the local strength reduction originated from the loss of effective load-bearing area, the respective contributions of surface geometric loss and subsurface microstructural damage could not be quantitatively separated. Third, the local compression tests characterized the short-term mechanical response of small regions corresponding to the groove wall surface, whereas the longer-term biomechanical consequences of such damage for surgically prepared, implant-relevant bone surfaces were not directly evaluated. Future studies should therefore examine fatigue-related mechanical deterioration under cyclic physiological loading across different bone conditions and seek to clarify the relative roles of surface geometry and subsurface microstructural damage.

## References

1. Kluess D, Bergschmidt P, Mueller I, Mittelmeier W, Bader R. Influence of the distal femoral resection angle on the principal stresses in ceramic total knee components. Knee. 2012;19:846–50. https://doi.org/10.1016/j.knee.2012.03.014
2. Ehlinger M, Ollivier M, Course S, Guerin A, Lantz É, Zahraa D, Bonnomet F, Bahlouli N. Effect of saw blade geometry on crack initiation and propagation on the lateral cortical hinge for HTO: finite element analysis. Orthop Traumatol Surg Res. 2019;105:1079–83. https://doi.org/10.1016/j.otsr.2019.04.026
3. James TP, Pearlman JJ, Saigal A. Predictive force model for haptic feedback in bone sawing. Med Eng Phys. 2013;35:1638–44. https://doi.org/10.1016/j.medengphy.2013.05.012
4. Demirtas A, Ural A. Material heterogeneity, microstructure, and microcracks demonstrate differential influence on crack initiation and propagation in cortical bone. Biomech Model Mechanobiol. 2018;17:1415–28. https://doi.org/10.1007/s10237-018-1035-6
5. Wang H, Satake U, Enomoto T. Reduction of sawing forces in bone cutting: innovative oscillating saw mechanism based on trajectory analysis. J Mater Process Technol. 2024;332:118563. https://doi.org/10.1016/j.jmatprotec.2024.118563
6. Wang H, Satake U, Enomoto T. A novel flexible-structured saw blade for bone cutting: reducing ploughing and promoting chip evacuation. J Mater Process Technol. 2025;344:119000. https://doi.org/10.1016/j.jmatprotec.2025.119000
7. Wazen RM, Currey JA, Guo H, Brunski JB, Helms JA, Nanci A. Micromotion-induced strain fields influence early stages of repair at bone–implant interfaces. Acta Biomater. 2013;9:6663–74. https://doi.org/10.1016/j.actbio.2013.01.014
8. Wang H, Satake U, Enomoto T. Modeling of oscillating bone sawing forces with instantaneous cutting speed and depth of cut. J Mater Process Technol. 2024;324:118225. https://doi.org/10.1016/j.jmatprotec.2023.118225
9. Shu L, Sugita N, Oshima M, Mitsuishi M. Design and experimental force analysis of a

novel elliptical vibration assisted orthopedic oscillating saw. Med Eng Phys. 2018;54:22–31. https://doi.org/10.1016/j.medengphy.2018.02.005

10. Shu L, Li S, Terashima M, Bai W, Hanami T, Hasegawa R, Sugita N. A novel self-centring drill bit design for low-trauma bone drilling. Int J Mach Tools Manuf. 2020;154:103568. https://doi.org/10.1016/j.ijmachtools.2020.103568
11. Robles-Linares JA, Axinte DA, Liao Z, Gameros A. Machining-induced thermal damage in cortical bone: necrosis and micro-mechanical integrity. Mater Des. 2021;197:109215. https://doi.org/10.1016/j.matdes.2020.109215
12. Chen J, An Q, Zou F, Yu D, Chen M. Analysis of low-frequency vibration-assisted bone drilling in reducing thermal injury. Mater Manuf Process. 2021;36:27–38. https://doi.org/10.1080/10426914.2020.1813891
13. Tawy GF, Rowe PJ, Riches PE. Thermal damage done to bone by burring and sawing with and without irrigation in knee arthroplasty. J Arthroplasty. 2016;31:1102–8. https://doi.org/10.1016/j.arth.2015.11.002
14. Chen YC, Tsai YJ, Hsiao HY, Chiu YW, Hong YY, Tu YK, Hsiao CK. Assessment of thermal osteonecrosis during bone drilling using a three-dimensional finite element model. Bioengineering. 2024;11:592. https://doi.org/10.3390/bioengineering11060592
15. Yeager C, Nazari A, Arola D. Machining of cortical bone: surface texture, surface integrity and cutting forces. Mach Sci Technol. 2008;12:100–18. https://doi.org/10.1080/10910340801890961
16. Liao Z, Axinte DA, Gao D. A novel cutting tool design to avoid surface damage in bone machining. Int J Mach Tools Manuf. 2017;116:52–9. https://doi.org/10.1016/j.ijmachtools.2017.01.003
17. Toksvig-Larsen S, Kroon PO, Ryd L. Improved bone cutting using a semirotating saw. A cadaver study of the cut surface on tibial condyles. Acta Orthop Scand. 1994;65:412–4. https://doi.org/10.3109/17453679408995481
18. Robles-Linares JA, Winter K, Liao Z. The effect of laser ablation pulse width and feed speed on necrosis and surface damage of cortical bone. Chin J Mech Eng. 2022;35:68.

https://doi.org/10.1186/s10033-022-00740-1

19. Arnold M, Zhao S, Ma S, Giuliani F, Hansen U, Cobb JP, Abel RL, Boughton O. Microindentation - A tool for measuring cortical bone stiffness? A systematic review. Bone Joint Res. 2017;6:542–9. https://doi.org/10.1302/2046-3758.69.BJR-2016-0317.R2
20. Feng L, Chittenden M, Schirer J, Dickinson M, Jasiuk I. Mechanical properties of porcine femoral cortical bone measured by nanoindentation. J Biomech. 2012;45:1775–82. https://doi.org/10.1016/j.jbiomech.2012.05.001
21. Kalidindi SR, Abusafieh A, El-Danaf E. Accurate characterization of machine compliance for simple compression testing. Exp Mech. 1997;37:210–5. https://doi.org/10.1007/bf02317861
22. Zhao S, Arnold M, Abel RL, Cobb JP, Ma S, Hansen U, Boughton O. Standardizing compression testing for measuring the stiffness of human bone. Bone Joint Res. 2018;7:524–38. https://doi.org/10.1302/2046-3758.78.BJR-2018-0025.R1
23. Turner CH, Burr DB. Basic biomechanical measurements of bone: A tutorial. Bone. 1993;14:595–608. https://doi.org/10.1016/8756-3282(93)90081-K
24. Peloquin JM, Santare MH, Elliott DM. Advances in quantification of meniscus tensile mechanics including nonlinearity, yield, and failure. J Biomech Eng. 2016;138:0210021. https://doi.org/10.1115/1.4032354
25. Lloyd EM, Hepburn MS, Li J, Mowla A, Hwang Y, Choi YS, Grounds MD, Kennedy BF. Three-dimensional mechanical characterization of murine skeletal muscle using quantitative micro-elastography. Biomed Opt Express. 2022;13:5879–99. https://doi.org/10.1364/boe.471062
26. Burr DB, Turner CH, Naick P, Forwood MR, Ambrosius W, Hasan MS, Pidaparti R. Does microdamage accumulation affect the mechanical properties of bone? J Biomech. 1998;31:337–45. https://doi.org/10.1016/S0021-9290(98)00016-5
27. Zimmermann EA, Ritchie RO. Bone as a structural material. Adv Healthc Mater. 2015;4:1287–304. https://doi.org/10.1002/adhm.201500070
28. Seref-Ferlengez Z, Kennedy OD, Schaffler MB. Bone microdamage, remodeling and

bone fragility: how much damage is too much damage? Bonekey Rep. 2015;4:644. https://doi.org/10.1038/bonekey.2015.11

29. Fletcher HH, Fisher MG. Surgical saw blade. United States patent US 6,503,253. 2003 Jan 7. https://patents.google.com/patent/US6503253B1/en

30. Vaidya SV, Dash KK. Tips and pearls: saw technique in total knee arthroplasty. In: Vaidya SV, Dash KK, editors. Basics in Hip and Knee Arthroplasty. New Delhi: Elsevier; 2017. pp. 222–9. https://shop.elsevier.com/books/basics-in-hip-and-knee-arthroplasty/vaidya/978-81-312-4888-1

31. Herregodts S, Verhaeghe M, Paridaens R, Herregodts J, Vermue H, Arnout N, De Baets P, Victor J. Soft-tissue penetration of the oscillating saw during tibial resection in total knee arthroplasty. Bone Joint J. 2020;102-B:1324–30. https://doi.org/10.1302/0301-620X.102B10.BJJ-2019-1602.R2

32. Ramos E, Grümberg U, Gutierrez J, Pezza V. Breakage of oscillating saw during intraoral vertical ramus osteotomy - a case report. Ann Maxillofac Surg. 2025;15:118. https://doi.org/10.4103/ams.ams_155_24

33. Tabariai E, Alexander G. An approach to retrieve a broken saw blade in an intraoral vertical ramus mandibular osteotomy. J Oral Maxillofac Surg. 2008;66:2412–3. https://doi.org/10.1016/j.joms.2007.12.043